\documentclass[amsmath,amssymb,11pt]{article}

\usepackage[T1]{fontenc}
\usepackage[latin9]{inputenc}

\usepackage{amstext}

\usepackage{hyperref}

\usepackage{graphicx}

\usepackage{esint}

\usepackage[english]{babel}
\usepackage[letterpaper]{geometry}
\begin{document}

\section*{Effective two-dimensional model does not account for geometry sensing by self-organized proteins patterns - Supplementary document}
{\bf Authors:} Jacob Halatek and Erwin Frey\\
{\bf Affiliation:} Arnold Sommerfeld Center for Theoretical Physics and Center for NanoScience, Department of Physics, Ludwig-Maximilians-Universit\"at M\"unchen, Theresienstra{\ss}e 37, D-80333 M\"unchen, Germany\\
{\bf Subject article:} Schweizer, J., Loose, M., Bonny, M., Kruse, K., M\"onch, I., and Schwille, P., 
Geometry sensing by self-organized protein patterns, 
{\em Proc. Natl. Acad. Sci. USA}, {\bf109}, 15283-15288 (2012)\\

\section*{Introduction} 
\label{sub:introduction}
Here we provide a thorough discussion of the model for Min protein dynamics proposed by Schweizer et al. \cite{Schweizer:2012ba}. The manuscript serves as supplementary document for our letter to the editor to appear in PNAS. Our analysis is based on the original COMSOL simulation files that were used for the publication. We show that all computational data in Schweizer et al. rely on exploitation of simulation artifacts and various unmentioned modifications of model parameters that strikingly contradict the experimental setup and experimental data. We find that the model neither accounts for MinE membrane interactions nor for any observed MinDE protein patterns. All conclusions drawn from the computational model are void. There is no evidence at all that persistent MinE membrane binding has any role in geometry sensing.

\section*{Summary of results and conclusions} 
\label{sec:conclusion}
\begin{itemize}
	\item The authors do not use the same parameters that are given in the article but a lower MinE/MinD ratio that also deviates from the experimental value.
	\item With the experimental MinE/MinD ratio no patterns emerge. (cf. Figure~\ref{fig:figures_patchesE19})
	\item With the lower MinE/MinD value patterns emerge but do not resemble the experiments nor the computational data presented in the article. (cf. Figure~\ref{fig:figures_rectpatchesE13a05}/\ref{fig:figures_rectpatchesE13a02a1})
	\item The quantitative data presented in Figure 5B/C/D in the paper cannot be reproduced by the proposed model (cf. Figure~\ref{fig:figures_Lshape}B and section \ref{ssub:alignment_to_curved_membrane_patches}).
	\item Alignment to the aspect ratio (Figure 5D in the paper) solely relies on self-coupling via the periodic boundaries in horizontal and vertical directions. (cf. Fig. 2/4/5)
	\item Alignment to curved membranes (Figures 5A/S7 in the paper) fails for gold layer sizes as used in the experiment. (cf. Figure~\ref{fig:figures_Lshape}D)
	\item Transient MinE membrane binding is an order of magnitude stronger than in the experiments. (cf. section \ref{ssub:residence_times_of_min_proteins})
	\item Adjusting MinE membrane binding to meet the experiment leads to loss of any patterns. (cf. section \ref{ssub:linear_stability_for_varying_mine_dynamics})
	\item The model assumes a very small bulk volume and is highly sensitive to volume effects in contrast to the experimental evidence and the claim in the paper. (cf. section \ref{ssub:total_particle_numbers_bulk_membrane_ratio_and_effective_2d_modeling})
\end{itemize}

We investigated the proposed model by means of linear stability analysis and numerical simulations. First, we note that the actual simulations provided to us by the authors use reduced MinE/MinD ratios ($C_{D0}/C_{E0}=2.23$ and $C_{D0}/C_{E0}=4.03$) that deviate from the experiments ([MinD]/[MinE]=1.6) and the published parameter values ($C_{D0}/C_{E0}=1.56$). For the experimental/published values we find pattern formation to be restricted to very small ratios of gold layer to membrane, or equivalently bulk volume to membrane. The model does not account for cytosolic volume explicitly, but the choice of the total protein densities indicates an effective bulk height below $6\mu m$. We find that rescaling the effective bulk volume by a small factor $\mathcal O(1)$ or explicitly increasing gold layer size yields loss of instability.
Hence, the model behavior described in the published simulations is limited to system sizes that deviate from the experiment by several orders of magnitude.  Moreover, in striking contradiction to the accompanying experiments and to the claim in the article, bulk size does have severe effects on protein patterns. The authors have compensated for the effects of reservoir size by adjusting intrinsic system parameters (total protein densities). This was not mentioned in the published article. It should go without saying that the need to adjust genuinely intrinsic system properties to keep certain desired phenomena invariant to variations of system size clearly proves that those phenomena are not intrinsic to the system. This directly contradicts the main experimental findings the model claims to account for. \\

However, even with these adjustments in place the model relies on employing simulation artifacts to reproduce the published data. We find that alignment to the aspect ratio (Fig. 5D in the paper) strictly requires periodic boundary conditions at the outer boundary of the gold layer. These cross-boundary couplings in horizontal and vertical directions controls the alignment angle, while the aspect ratio of the patch has a negligible effect on alignment. Wave alignment ceases and waves become disordered propagating blobs if gold layer sizes are inclreased to match the experimental setup or if cross-boundary coupling is disabled by replacing periodic boundary conditions with no-flux conditions. Plainly put, for simulations to resemble the presented set of computational data several distinct and independent artifacts and unphysical parameter adjustments have to be employed for each dataset individually. We were unable to determine the specific combinations of gold layer size, cross-boundary coupling, and total [MinE]/[MinD] density ratio, that yield the published data. Altogether, this invalidates the model on a conceptual level. \\

The model is claimed to extend and supersede all previous models by incorporating recent experimental evidence \cite{Park2011,Loose2011a} regarding MinE membrane interactions. We note that MinE membrane binding was already considered in the computational model by Arjunan and Tomita \cite{Arjunan2009}. Furthermore, the model contradicts the cited experimental references \cite{Park2011,Loose2011a} in several implications regarding transient MinE membrane binding.

Park et al. \cite{Park2011} have shown that unmasking the anti-MinCD domains in $\text{MinE}^{F7E/I24N}$ restores the wild type phenotype without membrane binding. We find that the model loses instability if the MinE membrane affinity is reduced. In contrast to the experiment the instability cannot be restored by any adjustment of the MinE recruitment rate (representing the unmasking of anti-MinCD domains). Hence, without any experimental support the model actually implies that MinE membrane binding is required for pattern formation in the first place. The claim that MinE membrane binding is supposed to be responsible for geometry sensing in particular is thereby unsubstantiated.

By means of the ratio of MinE/MinD residence times the relative strength of MinE membrane binding can be quantified. The individual residence times have been determined experimentally by Loose et al.\cite{Loose2011a}. We find that the value in the computational model exceeds the experimental value by one order of magnitude. While experiments show that the MinE membrane desity is always lower than the MinD density \cite{Loose2011a,Loose:2008ca}, the waves in the computational model contain up to ten times more MinE than MinD. In particular, we note that the computational data in Figure 5C cannot be reproduced. The simulations yield a MinE/MinD density ratio which is increased up to 16-fold compared to the published computational data in Figure 5C. This represents a 23-fold deviation from the experiments cited alongside \cite{Loose2011a}. The fact that the model assumes wave propagation based on very high concentrations of membrane bound MinE not co-localized with MinD invalidates the model on a qualitative level in addition to the various aforementioned quantitative discrepancies. We conclude that the model neither accounts for MinE membrane interactions nor for any observed MinDE protein patterns. Therefore, all conclusions drawn from the computational model are void.


\newpage
\tableofcontents
\section{Simulations} 
\label{sec:simulations}

\begin{figure}[f]
	\centering
		\includegraphics[width=\textwidth]{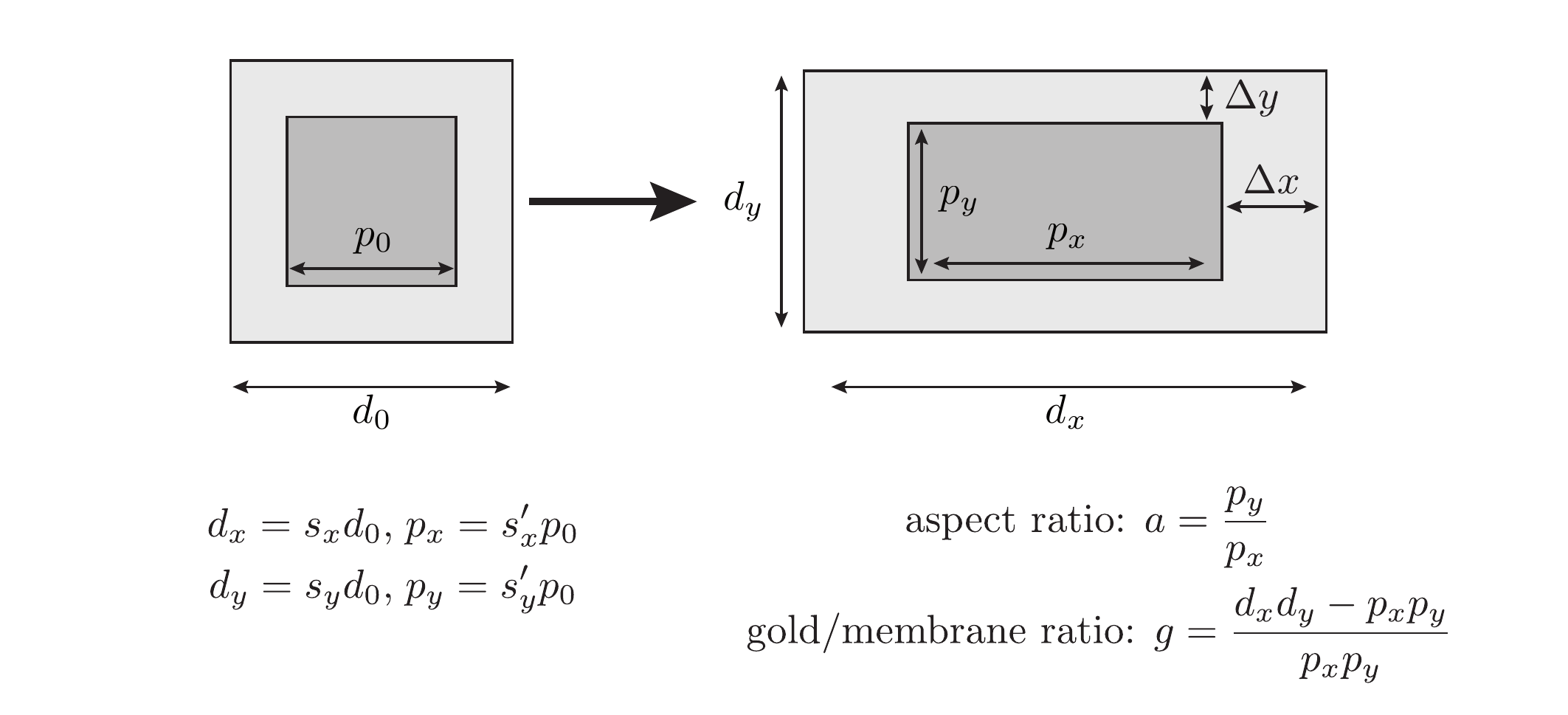}
	\caption{Model geometry and scaling operations.}
	\label{fig:figures_tech}
\end{figure}

\subsection{Model files and definitions} 
\label{sub:model_definition}
Since we were unable to reproduce the published results we asked the authors, following PNAS journal policies, to provide us with the full set of model files that were used to produce the data in the paper and accompanying supplement. The files we received are listed in Table 1 with the corresponding parameter configurations. In contrast to the value given in the paper $(C_{E0}=1.9 \cdot 10^3 /\mu m^2)$ the MinE density is set to $C_{E0}=1.3 \cdot 10^3 /\mu m^2$ in all simulations except the one for the L-shaped membrane patch (c.f. Fig 5 in the paper), where it is set to $C_{E0}=0.72 \cdot 10^3 /\mu m^2$. We did not receive an explanation or justification for the use of two different parameter values. Because the modification creates a conflict with the [MinD]/[MinE] ratio in the experiment\footnote{In the experiment a [MinD]/[MinE] ratio of 1.6 is maintained $(0.8\mu M \text{ MinD}, 0.5\mu M \text{ MinE})$. The original parameters agree with that $(2.9/1.9\approx 1.53)$, but the modified parameter do not ($2.9/1.3\approx 2.23$ and $2.9/0.72\approx 4.03$).},
we will consider both cases in the investigation of the computational model. Table 1 lists the protein densities, and the ratios of gold layer and membrane area used in the simulations. One notices that the gold layer area is always smaller or of comparable size as the membrane area but never very large as the experiments suggest. It is largest for the L-shape simulation where the MinE density is smallest. We further note that the boundary conditions are periodic. In combination with the small gold layer size a coupling of patch dynamics across the periodic boundaries appears likely. This will be considered below in the context of wave alignment to rectangular membrane patches. The aspect ratio in the corresponding model file is preset to $a=0.5$. The geometry is constructed by applying scaling operations on two squares with sides $d_0$ and $p_0$. The scaling operations are defined in Figure 1. All simulations of rectangular patches with altered aspect ratio or increased gold layer size are derived from this model file. For all simulations we have computed the solutions for $1.6 \cdot 10^4s$ from onset and compared the results with the solution at $1.2 \cdot 10^4s$ to ensure that the system reached a steady state. In simulations with increased gold layer sizes we have coarsened the gold layer mesh with increasing distance from the membrane patch. In these cases the mesh size in the gold layer was increased up to 5-fold of the value of the membrane mesh. We have carefully verified that this does not affect the patch dynamics by comparing the results with the simulations that employ a constant mesh size in the gold layer that corresponds to the membrane mesh.  In particular, we have compared the onset of pattern formation (first $1000s$) for original and coarse gold layer mesh. In addition we have verified that the final pattern after $1.6 \cdot 10^4s$ does not change if the simulation is continued for $1000s$ with a constant mesh in the gold layer that equals the membrane mesh.
\begin{table}[h]
	\centering
\begin{tabular}{|c||c|c|c|}
\hline
filename & $C_{D0}$ $[\mu m^{-2}]$ & $C_{E0}$ $[\mu m^{-2}]$ & $g$\\
\hline
\hline
AspectRatio\_Paper.mph & 2.9 & 1.3 & 0.56\\
\hline
circle\_S7.mph & 2.9 & 1.3 & 4.13\\
\hline
CouplingD15\_S11.mph & 2.9 & 1.3 & 0.38\\
\hline
CouplingD50\_S11.mph & 2.9 & 1.3 & 0.38\\
\hline
L\_shape\_Paper.mph & 2.9 & 0.72 & 6.68\\
\hline
largeGaps\_S10.mph & 2.9 & 1.3 & 1.67\\
\hline
LargePatch\_S09.mph & 2.9 & 1.3 & 0.35\\
\hline
note\_S7.mph & 2.9 & 1.3 & 4.14\\
\hline
serpentine\_S7.mph & 2.9 & 1.3 & 4.16\\
\hline
smallGaps\_S10.mph & 2.9 & 1.3 & 0.61\\
\hline
smallPatch\_S9.mph & 2.9 & 1.3 & 0.86\\
\hline
\end{tabular}
\caption{Comsol Multiphysics simulation files provided by the authors. We list the preset values for the total MinD and MinE densities, $C_{D0}$ and $C_{E0}$, as well as the gold/membrane ratio $g$ (Fig. 1).}
\end{table}

\subsection{Simulations with $C_{E0}=1.9 \cdot 10^3 /\mu m^2$ $(\text{[MinD]}/\text{[MinE]}=1.53)$} 
\label{sub:original_parameters}
In all following simulations the total MinE density is set to $C_{E0}=1.9 \cdot 10^3 /\mu m^2$ unless noted otherwise.
\subsubsection{Alignment to rectangular membrane patches} 
\label{ssub:alignment_to_rectangular_membrane_patches}
\begin{figure}[t]
	\centering
		\includegraphics[width=\textwidth]{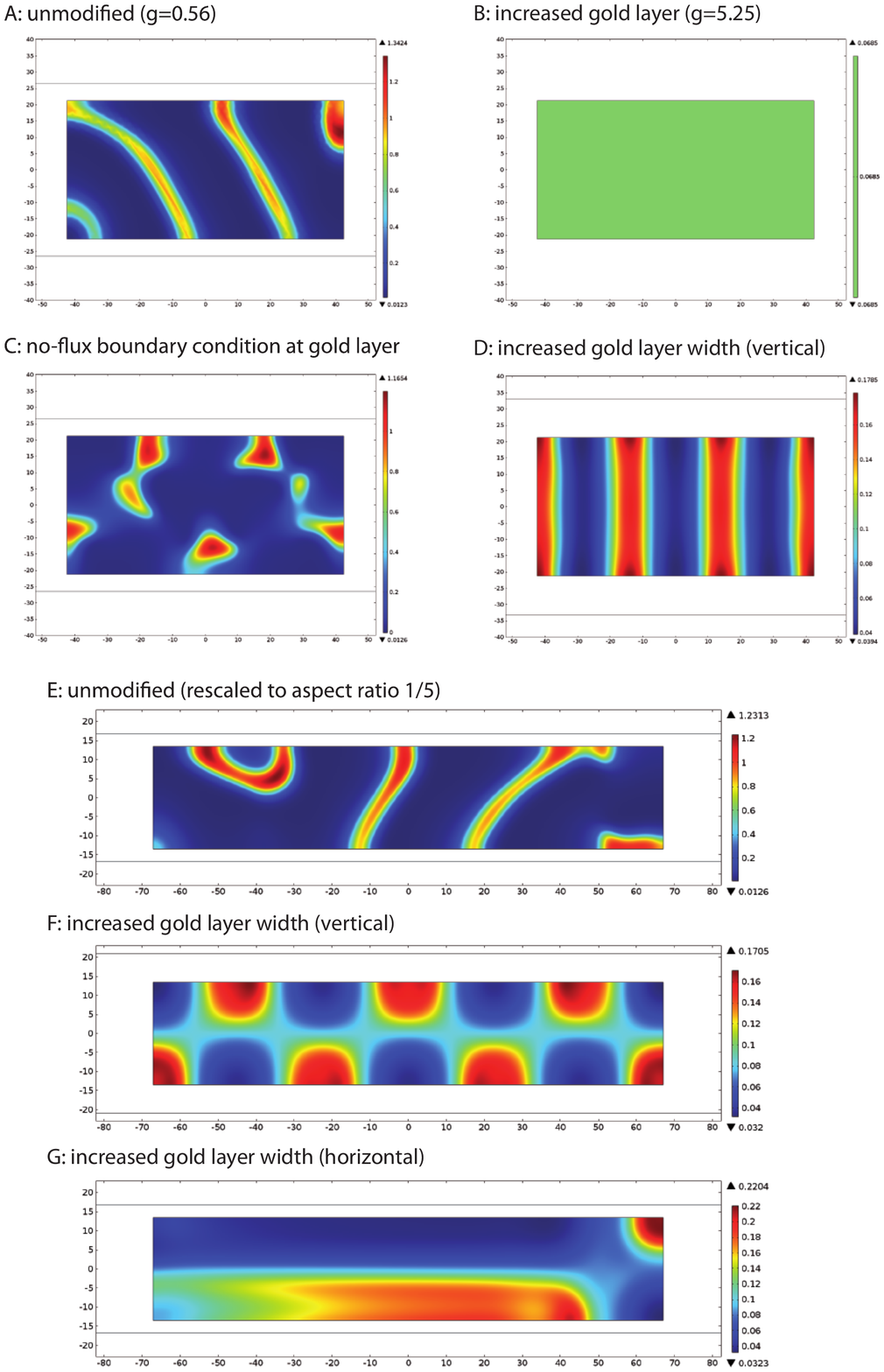}
	\caption{Dynamics on rectangular patches for $C_{E0}=1.9 \cdot 10^3 /\mu m^2$ as published in the paper. All simulations are based on the file AspectRatio\_Paper.mph}
	\label{fig:figures_rectpatchesE19}
\end{figure}

The simulation file corresponding to Figure 5D in the original paper (AspectRatio\_Paper.mph) produces the reported alignment to the diagonal  for the preset aspect ratio $a=0.5$ (c.f. Figure~\ref{fig:figures_rectpatchesE19}A).
We started out increasing the size of the surrounding gold layer ($2d_0$, $g=5.25$) and found that the system settles in an homogeneous stable state (Figure~\ref{fig:figures_rectpatchesE19}B).
This implies that pattern formation requires a very small gold layer to membrane ratio. Next, we went to investigate wave coupling across periodic boundaries and the impact on wave alignment in the original model simulations. To eliminate any possible coupling across periodic boundaries we replaced the periodic condition by a no-flux condition. As a result waves ceased to align to the diagonal in contrast to the data in the original paper. Instead we observed a disordered pattern of oscillatory blots without any apparent alignment but with preference for the edges and center of the patch (Figure~\ref{fig:figures_rectpatchesE19}C). This leads us to the conclusion that the alignment observed in the simulation must result from the coupling across periodic boundaries. To shed light on this matter we changed the aspect ratio of the surrounding gold layer, leaving the aspect ratio and size of the membrane patch unaltered. 
Increasing either the horizontal ($x$) or vertical ($y$) width of the gold layer by a factor 1.5 or more resulted in a stable homogeneous state without any pattern on the membrane patches. Increasing the width merely by a factor 1.25 led to a standing wave pattern aligned to the major axis of the patch (Figure~\ref{fig:figures_rectpatchesE19}D). This complex change in dynamics due to a small perturbation of the surrounding geometry strongly indicates that the ratio of gold layer area to membrane area as well as the anisotropic coupling across boundaries mainly regulate the formation and selection of patterns on the rectangular membrane patches. Since the alignment angle is regulated by the aspect ratio of the patch in the experiment one might expect a similar result in the simulation. The paper clearly states that the model accounts for these experimental findings. In our attempts to reproduce this claim we rescaled the membrane patch along with the surrounding gold layer in order to obtain a membrane patch and gold layer with an aspect ratio of 0.2. For this case the paper reports alignment of waves to the long axis in the experiment as well as in the simulation. In contrast, we find that waves start at the edges of the patch and align to the diagonals in the initial phase. After $1.6 \cdot 10^4s$ the system consists of a mixture of disordered waves and imperfect drifting spirals with no alignment to a specific axis (Figure~\ref{fig:figures_rectpatchesE19}E). Again, slightly increasing the size of surrounding gold layer leads to a complete loss of any patterns and a spatially uniform density. Increasing the vertical width by a factor 1.25 leads to a pattern of standing nodes attached to the edges of the long axis with odd symmetry with respect to the short axis (Figure~\ref{fig:figures_rectpatchesE19}F). Increasing the horizontal width instead leads to domains forming consecutively in the upper and lower half of the patch (Figure~\ref{fig:figures_rectpatchesE19}G). 

To conclude, we find that the published data (Figure 5D in the paper) cannot be recovered. Waves do not align to the aspect ratio of the membrane patch. On the contrary, alignment arises as pure result of the cross-boundary coupling, and only if the surrounding gold layer area is very small compared to the patch size. Hence, alignment is not intrinsic to the membrane geometry as the authors claim but, in fact, the exact opposite: there is no intrinsic alignment (i.e. independent of cross-boundary coupling) to the membrane geometry at all.
This point is further emphasized by the simulation that the authors used to investigate decoupling of patters due to increased patch distance (large\_gaps\_S10.mph). In contrast to the published data (Figure S10B in the paper), we find that the system settles in a homogeneous stationary state (Figure~\ref{fig:figures_patchesE19}A) for the published parameter set.
%
%
\subsubsection{Alignment to curved membrane patches} 
\label{ssub:alignment_to_curved_membrane_patches}
\begin{figure}[t]
	\centering
		\includegraphics[width=\textwidth]{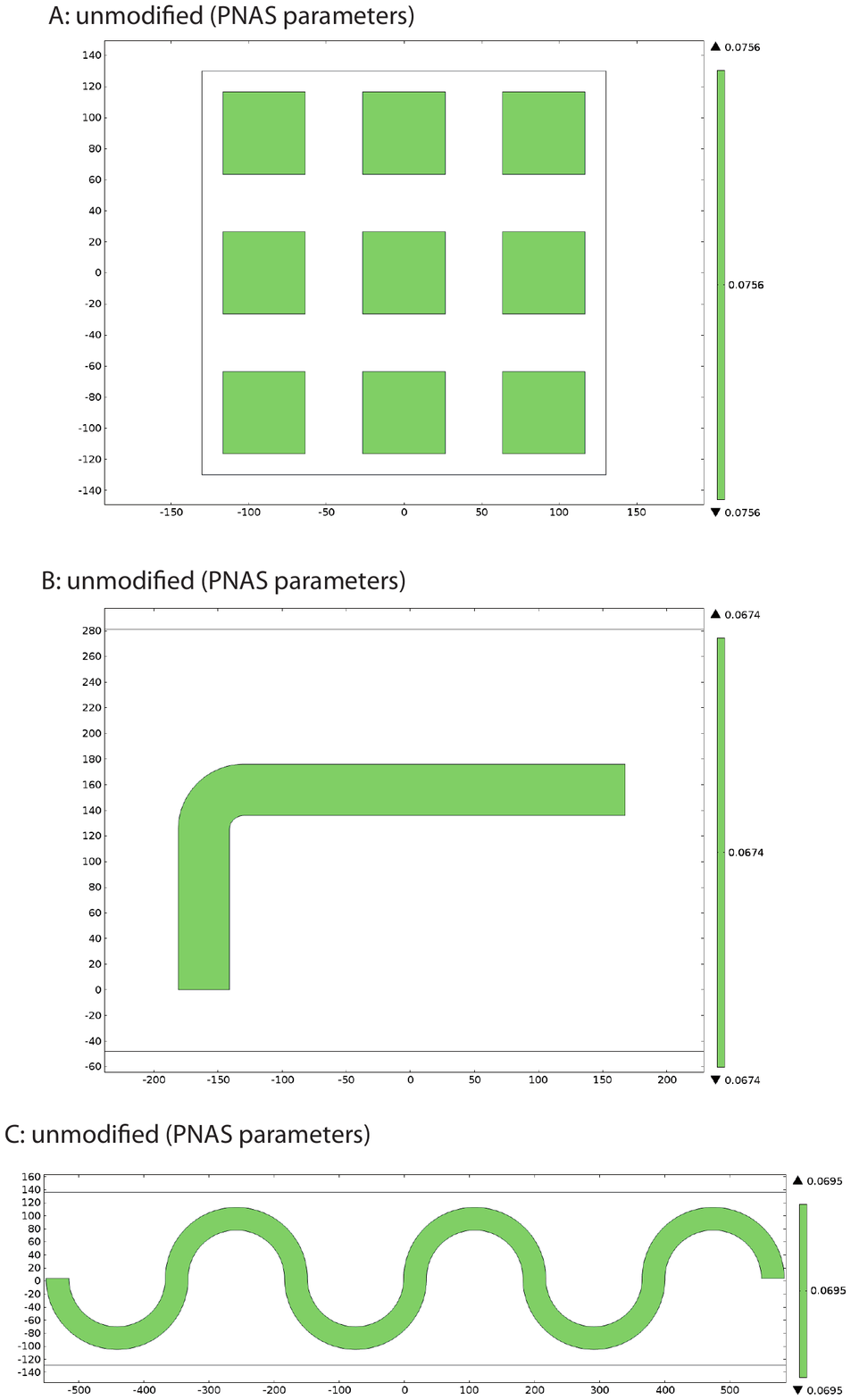}
	\caption{Absence of patterns on quadratic and curved patches for $C_{E0}=1.9 \cdot 10^3 /\mu m^2$ as published in the paper.}
	\label{fig:figures_patchesE19}
\end{figure}

The second quantitative dataset provided by the computational analysis in the paper concerns the alignment of Min protein waves to curved membrane patches. In the simulation files the membrane patches are embedded in rectangular domains representing surrounding gold layer. The preset ratio of gold layer area to membrane area $g$ can be found in Table 1. Again, we ran all simulations with the total MinE and MinD concentrations as published in the paper and found that the system settled in a stable homogeneous state (Figure~\ref{fig:figures_patchesE19}A-C). It is obvious that a rectangular domain surrounding curved patches cannot be made arbitrarily small to yield dynamical instabilities. Therefore, other parameters need to be changed to trigger any pattern formation process and enable comparison with the published results. To reproduce the published results the initial MinE density has been reduced by the authors to $C_{E0}=1.3 \cdot 10^3 /\mu m^2$ ($\text{[MinD]/[MinE]}=2.23$) in all simulations except the one for the L-shape simulation, in which case the initial MinE concentration had been set to $C_{E0}=0.72\cdot 10^3/ \mu m^2$, i.e. $\text{[MinD]/[MinE]}=4.03$. This does not come as a surprise, given that the sensitivity of Min protein patterns to [MinD]/[MinE] ratios is well documented in the literature \cite{Halatek:2012gm, Kerr2006, Huang:2003bc, Meacci2005}. Nonetheless, the necessity to change this parameter in this particular case directly contradicts the accompanying experiments the model strives to account for. To gather insight into the reported model dynamics we repeated the simulations with the reduced total MinE density that was preset in the simulations files. The results are discussed in the next section.



\subsection{Simulations with $C_{E0}=1.3 \cdot 10^3 /\mu m^2$ $(\text{[MinD]}/\text{[MinE]}=2.23)$} 
\label{sub:corrected_parameters}
In all following simulations the total MinE density is set to $C_{E0}=1.3 \cdot 10^3 /\mu m^2$ unless noted otherwise.

\subsubsection{Alignment to rectangular membrane patches} 
\label{ssub:alignment_to_rectangular_patches}
\begin{figure}[t]
	\centering
		\includegraphics[width=\textwidth]{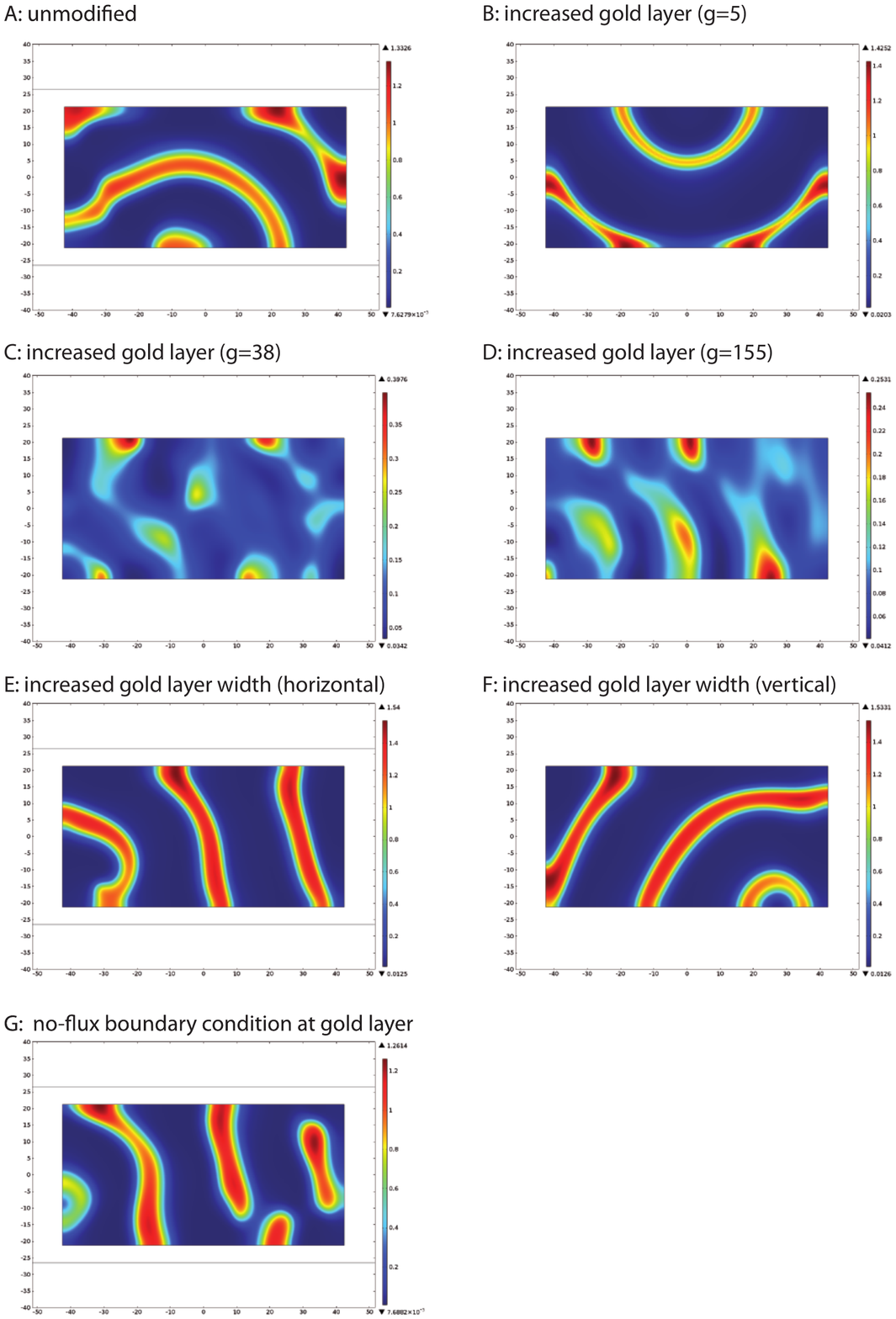}
	\caption{Dynamics on rectangular patches for $C_{E0}=1.3 \cdot 10^3 /\mu m^2$ as preset in the simulation files. All simulations are based on the file AspectRatio\_Paper.mph}
	\label{fig:figures_rectpatchesE13a05}
\end{figure}

Running the simulation of rectangular patches with the preset MinE concentration $C_{E0}=1.3 \cdot 10^3 /\mu m^2$ yields target patterns forming approximately at the center of one of the long edges. The target waves propagate along the short axis (Figure~\ref{fig:figures_rectpatchesE13a05}A) and the pattern remains disordered. Increasing the size of surrounding gold layer first stabilizes the target pattern and yields clean target waves centered at the edge of the long axis for $2d_0$ (Figure~\ref{fig:figures_rectpatchesE13a05}B). However, further increase of the gold layer size leads to a destabilization of the target pattern and we observe disordered propagating blots for $5d_0$ (Figure~\ref{fig:figures_rectpatchesE13a05}C) and disordered standing waves/blots for $10d_0$ (Figure~\ref{fig:figures_rectpatchesE13a05}D). These waves and blots are typically attached to the inner boundaries of the patch and tend to form and propagate rather chaotically within the patch domain. As none of these patterns resembles the reported data from the paper we investigated the effect of cross-boundary coupling. Increasing the width of the gold layer in either direction yields waves roughly aligned to the diagonal (Figure~\ref{fig:figures_rectpatchesE13a05}E/F). Leaving the gold layer unaltered but replacing the periodic boundary conditions with no-flux boundary conditions yields disordered waves and blots that are aligned to the long axis (Figure~\ref{fig:figures_rectpatchesE13a05}G). Once more, the results indicate that the reported alignment to the diagonal requires careful tuning of the cross-boundary coupling and gold layer size.

\begin{figure}[t]
	\centering
		\includegraphics[width=\textwidth]{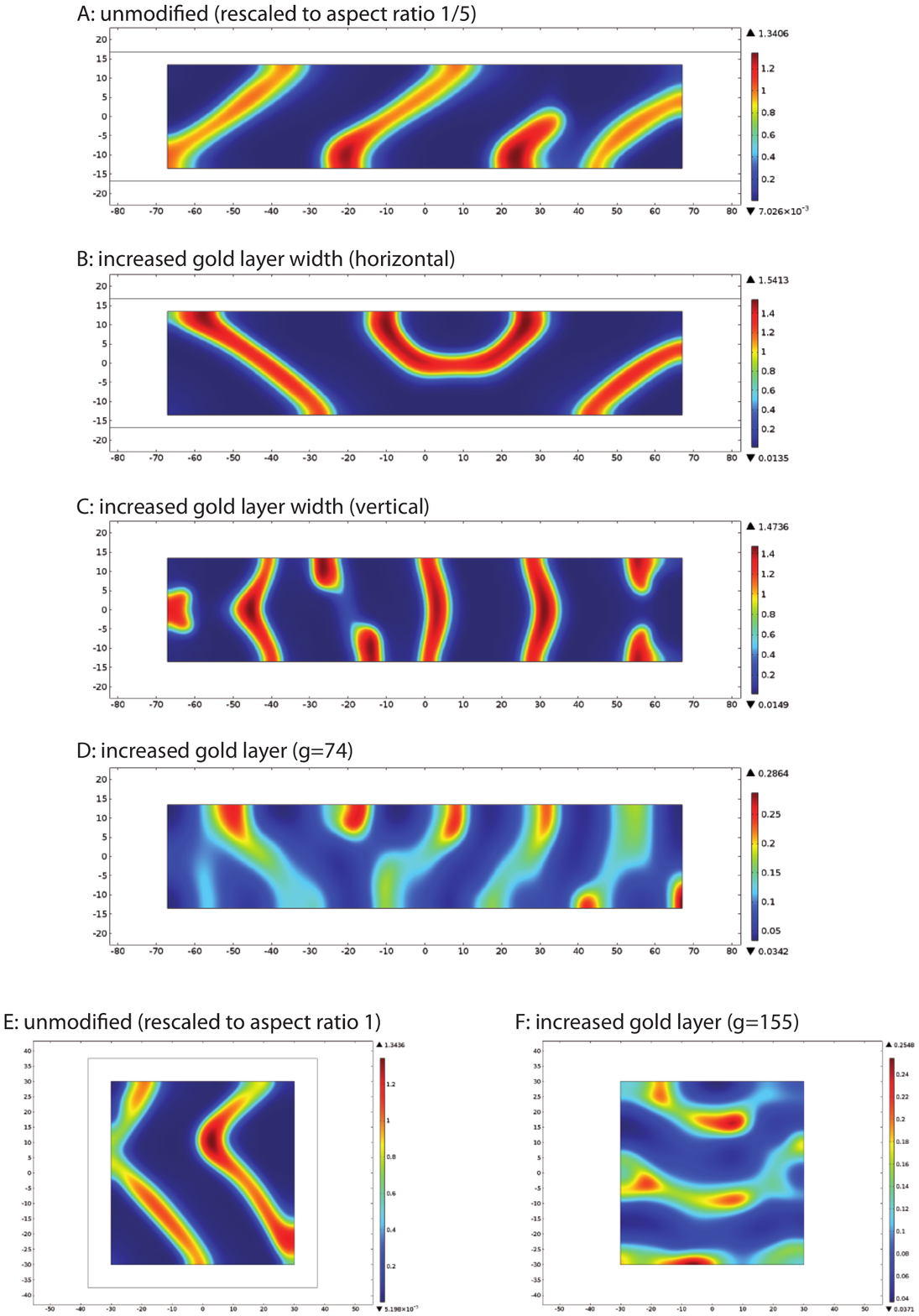}
	\caption{Dynamics on rectangular patches for $C_{E0}=1.3 \cdot 10^3 /\mu m^2$ as preset in the simulation files. All simulations are based on the file AspectRatio\_Paper.mph}
	\label{fig:figures_rectpatchesE13a02a1}
\end{figure}

To investigate the effect of the patch geometry we repeated the simulations for different aspect ratios of the membrane patch. Rescaling the whole system to obtain a membrane patch with aspect ratio 0.2 yields counter-propagating waves aligned to the diagonal and originating at two diagonally opposed corners (Figure~\ref{fig:figures_rectpatchesE13a02a1}A). Note that the paper reports alignment to the long axis for this aspect ratio in the experiment and corresponding simulations. Increasing the horizontal width of the gold layer yields diagonally aligned waves originating from horizontally opposed corners (Figure~\ref{fig:figures_rectpatchesE13a02a1}B). We find that the alignment to the long axis requires increasing the vertical width of the rescaled gold layer (Figure~\ref{fig:figures_rectpatchesE13a02a1}C). In this case wave originate from all four corners of the patch, and waves originating from vertically opposed corners merge to align to the long axis. However, the pattern is symmetric as counter-propagating waves annihilate each other. As in the previously discussed simulations for patches with aspect ratio $a=0.5$ we find disordered waves and blots attached to the membrane edges if the area of the gold layer is increased uniformly (Figure~\ref{fig:figures_rectpatchesE13a02a1}D). Similar results are obtained for membrane patches with aspect ratio 1 (Figure~\ref{fig:figures_rectpatchesE13a02a1}E/F). Hence, any realistically sized gold layer where cross-boundary coupling can be excluded results in a disordered state that does not resemble the experimental or computational data from the paper.
 
\begin{figure}[t]
	\centering
		\includegraphics[width=\textwidth]{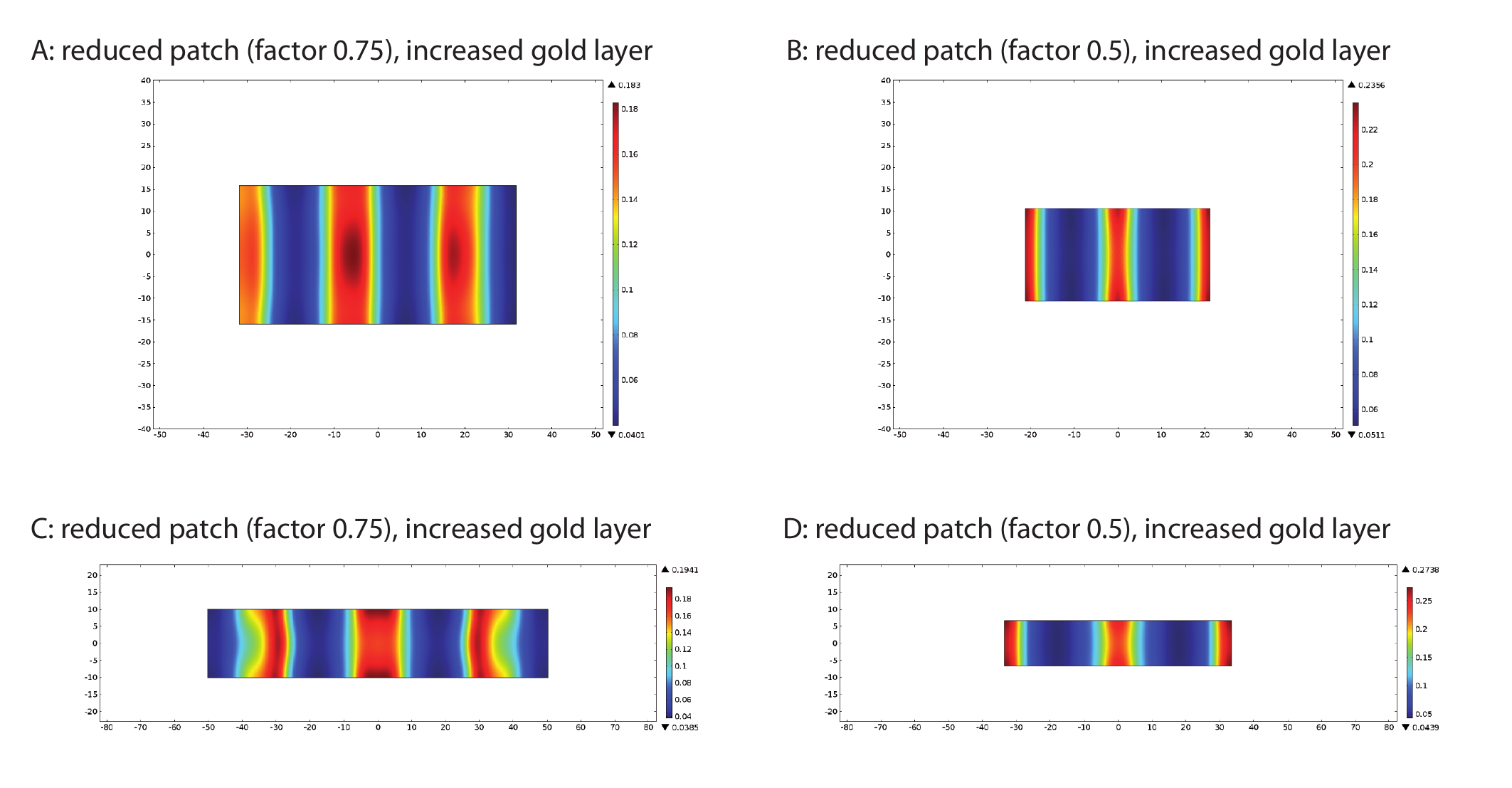}
	\caption{Dynamics on smaller rectangular patches for $C_{E0}=1.3 \cdot 10^3 /\mu m^2$. All simulations are based on the file AspectRatio\_Paper.mph}
	\label{fig:figures_rectpatchesE13small}
\end{figure}

As the experimental data for large membrane patches indicates a much broader distribution of alignment angles one might suspect that the failure of the model originates from the choice of patch size. In particular, the experiments imply that misalignment occurs if the patch is much larger than the wavelength of the patterns. Therefore, we investigated the model dynamics for large gold layers and smaller patches with constant aspect ratio. We find that reducing the patch size to $0.75p_0$ and $0.5p_0$ results in a pure standing wave pattern aligned to the long axis for aspect ratios $a=0.5$ (Figure~\ref{fig:figures_rectpatchesE13small}A/B) and $a=0.2$ (Figure~\ref{fig:figures_rectpatchesE13small}C/D). Hence, the patch size is not the cause for the disordered waves and failed alignment.

We conclude that a commensurable alignment can only be achieved by exploiting cross-boundary coupling effects and tuning the total MinE density. For instance, to recover alignment to the long axis one needs to employ an anisotropic rescaling of the cross-boundary coupling in y-direction and reduce the MinE density. This demonstrates that the [MinE]/[MinD] ratio sensitively regulates the alignment angle in presence of active cross-boundary coupling. Moreover, reducing the total MinE density entails that the dynamical instability is not lost when the gold layer is increased. This enables us to study the alignment to curved membrane patches in the following section.


\subsubsection{Alignment to curved membrane patches} 
\label{ssub:alignment_to_curved_membrane_patches}
\begin{figure}[t]
	\centering
		\includegraphics[width=\textwidth]{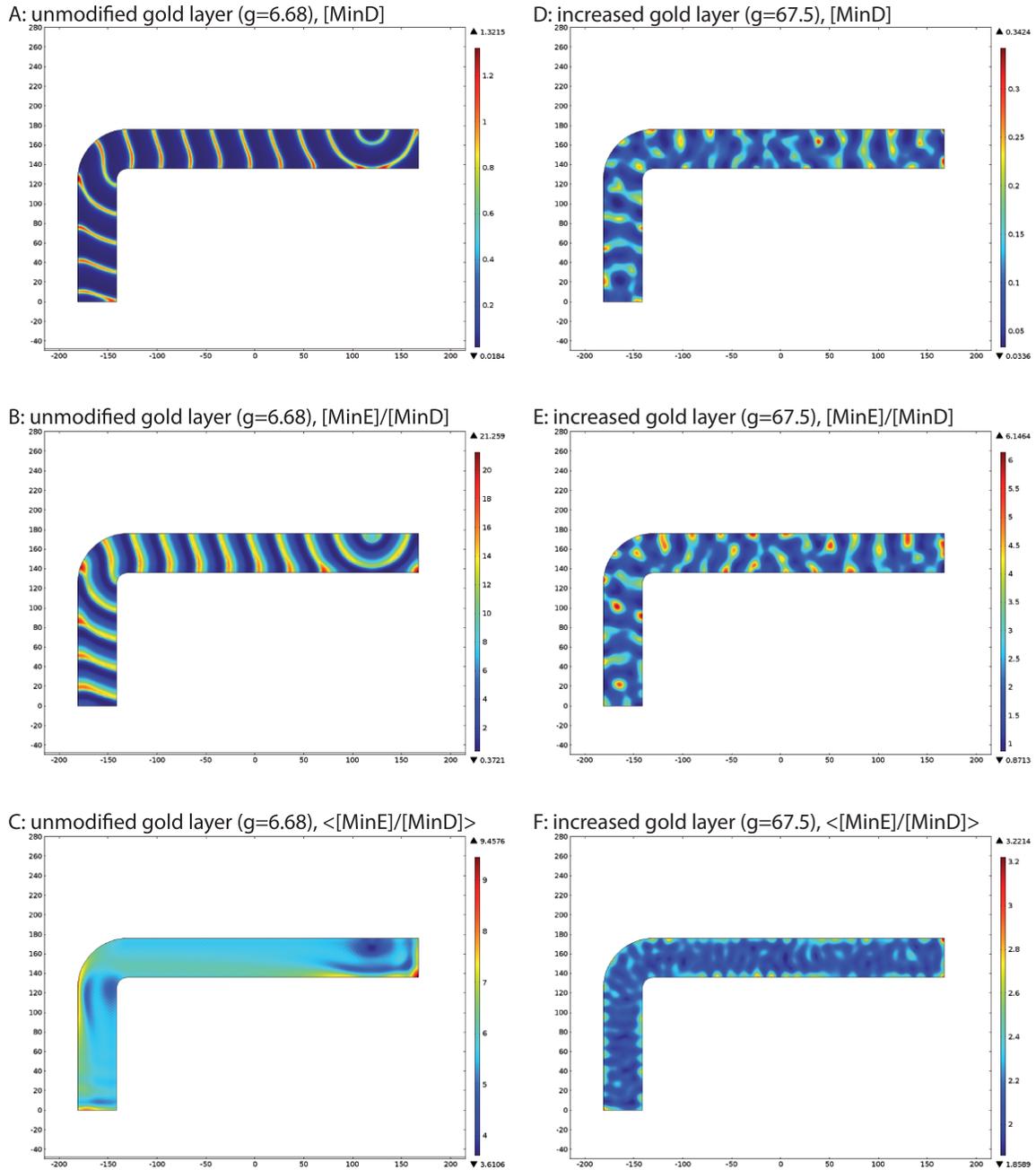}
	\caption{Dynamics on L-shaped patches for $C_{E0}=1.3 \cdot 10^3 /\mu m^2$. All simulations are based on the file L\_shape\_Paper.mph}
	\label{fig:figures_Lshape}
\end{figure}

As mentioned above the MinE concentration in the L-shape model file provided by the authors is preset to $C_{E0}=0.72 \cdot 10^3 /\mu m^2$ while it is set to $C_{E0}=1.3 \cdot 10^3 /\mu m^2$ in all other model files (cf. Table 1). For the sake of a systematic study we adjusted the MinE concentration to match all other simulations. With an otherwise unaltered model file we find that wave trains align to the patch as reported in the paper after about $10^4s$ (Figure~\ref{fig:figures_Lshape}A). While the phenomenology is quite similar, i.e. waves are aligned to the long rectangular sections then turn into the kink and realign afterwards, neither wave velocities nor [MinE]/[MinD] ratios along the patch match the data from the paper. The paper reports velocities in the range $1.3\mu m/s-3.2\mu m/s$ (on the rectangular section and at the outer edge of the curve) while we find waves with velocities in the range $0.8\mu m/s-2.2\mu m/s$. This shows that the data in the paper is not recovered with the current parameters. The ability of waves to realign is ascribed to the dynamics that locally increase the [MinE]/[MinD] ratio (and thereby the wave velocity) at the outer part of the curve. This mechanism for geometry sensing is ascribed to MinE membrane binding, quoting the paper:
\begin{quote}
	``When we plotted the ratio between the activator MinE and the membrane-bound ATPase MinD along the travel path of the membrane, we found that this ratio is significantly higher at the outer part compared to the inner part of the wave, thereby accelerating the detachment of the proteins from the membrane (23) (Fig. 5C). [...] Importantly, we could only reproduce this behavior when we considered transient binding of MinE to the membrane in our model.''
\end{quote}
In this context the authors cite their previous work \cite{Loose2011a} where the local [MinE]/[MinD] ratio within a wave had been quantified. These previous experiments revealed that the [MinE]/[MinD] membrane density ratio peaks at the rear of the wave. The maximal [MinE]/[MinD] ratio is about 0.9 and it marks the point where the protein flux off the membrane is maximized and drives wave propagation. For the computational model the authors report a [MinE]/[MinD] ratio about 1.33 at the outer part vs. 1.15 at the inner part, (cf. Figure 5C in the paper). 

\begin{figure}[t]
	\centering
		\includegraphics[width=\textwidth]{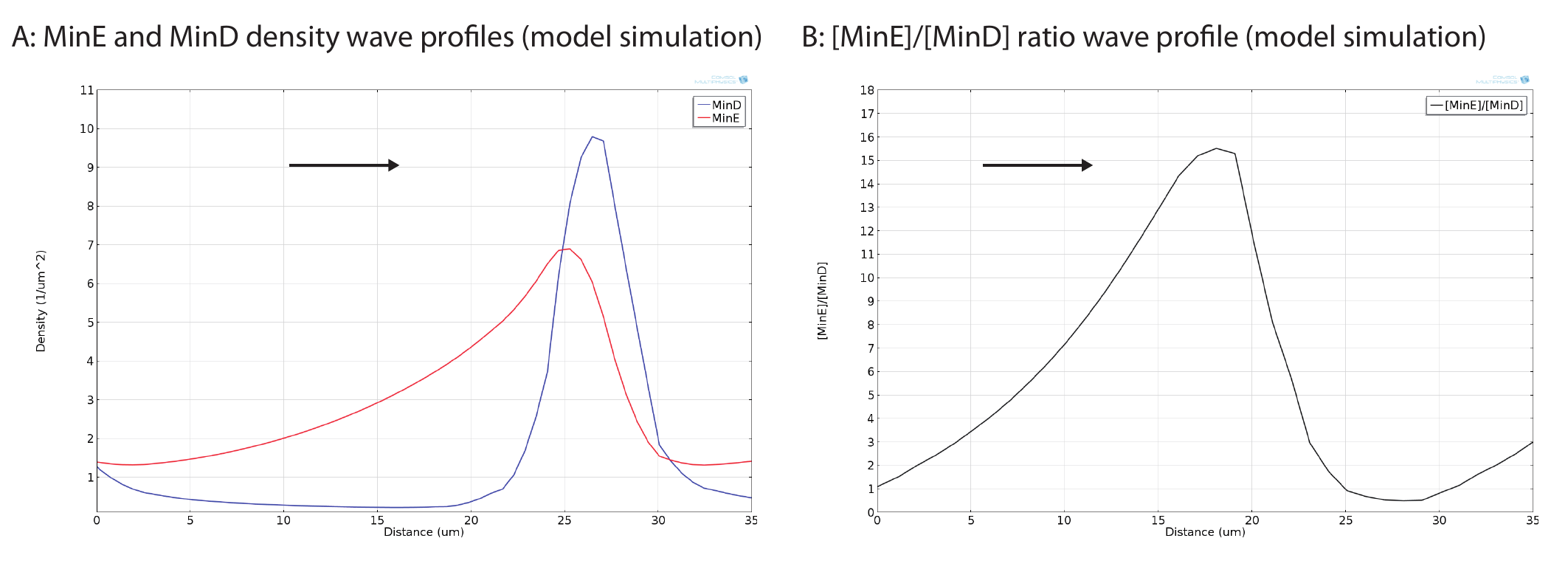}
	\caption{Computational data of protein density profiles and [MinE]/[MinD] ratios along a wave in the L-shape simulation.}
	\label{fig:figures_waveprofiles}
\end{figure}

We find that the simulations yield a local [MinE]/[MinD] ratio about 21 at the outer part of the curve and 15 at the inner part (Figure~\ref{fig:figures_Lshape}B). This represents a striking 16-fold deviation from the simulation data in the paper and a 23-fold deviation from the experimental data the authors cite in this particular context to support the model. We have also plotted the mean profile of the [MinE]/[MinD] density ratio on the patch in Figure~\ref{fig:figures_Lshape}C. Time-integration was performed over 950s which is the period of the envelope modulating the amplitude of the protein waves \cite{cross2009}. As expected the mean [MinE]/[MinD] profile shows a maximum at the outer part of the curve and a minimum at the inside. The mean [MinE]/[MinD] density ratio takes values between 3.6 and 9.5 throughout the patch.  Comparison of the wave profile for MinE and MinD densities and [MinE]/[MinD] ratios from the rectangular section of the L-shaped patches with experimental data published by the authors (see supplementary figure 1 in \cite{Loose2011a}) reveals a similar quantitative inconsistency, cf. Figure~\ref{fig:figures_waveprofiles}.
We note that these results are equally recovered if the simulations are performed with a further reduced total MinE concentration  $(C_{E0}=0.72 \cdot 10^3 /\mu m^2)$ that was preset in the model file we received from the authors. The most notable difference was that alignment was already established after about $1000s$ with $C_{E0}=0.72 \cdot 10^3 /\mu m^2$, hence, one order of magnitude earlier than with $C_{E0}=1.3 \cdot 10^3 /\mu m^2$. The origin of the quantitative discrepancies between [MinE]/[MinD] membrane density ratios in the published dataset and the simulations remains elusive. It appears to be intrinsic to the model as it relies on very strong MinE membrane binding. This will be discussed in the next section.

Before we go into that, we address the question if waves are sustained for realistically large gold layers as the experiments dictate. We note that the variation of the cytosolic protein densities is already very small at the system boundary (well below 1\%). Still, when the surrounding gold layer is increased ($g=72$) waves become disordered and cease to align (Figure~\ref{fig:figures_Lshape}D). This observation is consistent with the previous simulations of rectangular patches (cf. Figure~\ref{fig:figures_rectpatchesE13a05}C/D and \ref{fig:figures_rectpatchesE13a02a1}D/F). This observation emphasizes that coupling across the periodic boundary and the gold/membrane ratio $g$ are two distinct aspects of the geometry affecting the model dynamics. The irregular waves found for the increased gold layer do not span the patch width but are attached to the interior edges. Also, waves do not align and realign after passing through the curve. Apparently, transient MinE membrane binding does not facilitate the local increase of the [MinE]/[MinD] ratio at the outer part of the curve (Figure~\ref{fig:figures_Lshape}E/F) any more. 
To conclude: For simulations to resemble the data depicted in the paper, both, the total MinE density as well as the size of the surrounding gold layer need to be fine tuned. While this procedure already conflicts with the experiments in the first place, the simulations one recovers deviate from the published experimental and computational data by one order of magnitude in the very quantity that is argued to be key for the phenomenon of geometry sensing that the computational analysis tries to explain.


\section{Incompatibility of model assumptions and experimental data} 
\label{sec:compatibility_of_model_assumptions_with_experimental_data}
So far we have established by means of numerical simulations that the model does not account for any experimental observation in the paper. 
Clearly, this conflicts with the final conclusion reached by the authors in the paper's abstract:
\begin{quote}
	``Using a computational model we quantitatively analyzed our experimental findings and identified persistent binding of MinE to the membrane as requirement for the Min system to sense geometry. Our results give insight into the interplay between geometrical confinement and biochemical patterns emerging from a nonlinear reaction-diffusion system.''
\end{quote}
The molecular basis of the model is set by the authors' previous experimental work \cite{Loose2011a} and the work by Park et al. \cite{Park2011}. The main intention is to address the role of MinE membrane binding by means of the computational analysis, quoting from the paper:
\begin{quote}
	``Recently, two reports have shown that MinE persists at the MinD-membrane surface after activation of the MinD ATPase (23, 30). Although persistent binding of MinE appears to be important for its ability to completely remove MinD from the membrane (23), its possible role for the ability of the Min system to organize the interior of the cell has so far not been addressed. Our model extends previous models by incorporating that MinE transiently interacts with the membrane during the activation of MinD. This description gives a unified account of all currently known stable Min-protein patterns in vivo and in vitro as will be discussed in detail elsewhere.''
\end{quote}
The final statement is proven false by the results from the previous sections. In this context the question arises whether the model can be used for an assessment of MinE membrane binding at all. The research by Park et al. \cite{Park2011} strongly suggests that MinE membrane interactions take place. We want to emphasize that we do not question these experimental findings in any way. On the contrary we find these results very helpful to understand the molecular basis of MinE protein dynamics as reflected in our own research \cite{Halatek:2012gm}.

In this section we show that the model conflicts with the available experimental evidence in it's molecular basis.  Our analysis will show that the model operates in a kinetic regime that contradicts the experimental data. As such, the model is invalid as a theory for MinE membrane interactions. No conclusions about MinE membrane interaction can be drawn from it. To establish a mutual basis in terminology we start by describing how the proposed model introduces MinE membrane interactions. We discuss the distinct molecular processes the model is based on and compare the qualitative and quantitative implications  with experimental data. We limit the comparison to experimental research papers that are explicitly cited in the paper to motivate the model. In the last part we will address the role of particle numbers, bulk-membrane ratio, and the validity of an effective 2D modeling.

\subsection{The role of MinE membrane interactions} 
\label{ssub:the_role_of_mine_membrane_interactions}
The model assumes that cytosolic MinE is able to sense membrane bound MinD-ATP and form a MinDE complex upon recruitment. These two steps are introduced and quantified by the MinE recruitment rate $\omega_E=5\cdot 10^{-4}\mu m^2/s$. Upon MinDE complex formation, MinE stimulates MinD ATPase activity which leads to the detachment of MinD-ATP (nucleotide exchange is neglected in the model) from the membrane. The stimulation of MinD ATPase with MinD detachment is quantified by the sum of detachment rates $\omega_{de}\equiv\omega_{de,c}+\omega_{de,m}=0.88s^{-1}$. Persistent MinE binding is introduced by enabling the MinE dimer bound in a MinDE complex to directly interact with the membrane. Upon stimulation of MinD ATPase activity MinE remains bound to the membrane with a certain probability $p_m={\omega_{de,m}}/{\omega_{de}}=0.91$ or detaches instantly along with MinD with probability $p_c=1-p_m={\omega_{de,c}}/{\omega_{de}}=0.09$. Of course, within the notion of a deterministic model $p_m$ and $p_c$ can be interpreted as fraction of MinE concentration that remains membrane bound or becomes cytosolic upon stimulation of MinD ATPase activity, respectively. The strength of the MinE-membrane bond is characterized by the MinE detachment rate $\omega_e=0.08s^{-1}$. The parameter $\omega_{ed}=2.5\cdot 10^{-3}\mu m^2/s$ quantifies MinE-MinD reassociation at the membrane. All these parameters can be tuned individually to study the model dynamics. However, no experimental data exists to support any particular parameter choice. Moreover, mutations studies (e.g. regarding the MTS of MinE) are unlikely to affect only one corresponding model parameter alone.
To constrain MinE membrane binding and MinE-MinD interactions in the model we will take the experimentally determined Min protein residence times into account. This enables us to establish a direct relation between quantitative experimental data and the corresponding model parameters. 

\subsubsection{Residence times of Min proteins in vitro} 
\label{ssub:residence_times_of_min_proteins}
The residence $<\tau_D>$ and $<\tau_E>$ times of MinD and MinE along the protein wave have been quantified by the authors in previous experiments \cite{Loose2011a}.  The experiments revealed that MinE remains longer in a membrane bound state than MinD which indicates transient MinE membrane binding. While the individual residence times increase from front to rear of the wave, the ratio of residence times $\Delta\tau=<\tau_E>/<\tau_D>$ appears to be constant throughout the wave $(\Delta\tau_\text{exp}\approx1.31)$. Therefore, we can use the ratio of residence times as characteristic parameter to quantify transient MinE membrane interaction in context of any specific model.

The mean residence time of MinD $<\tau_D>$ is given by the time it takes cytosolic or membrane-bound MinE to sense and attach to a membrane-bound MinD, and the time MinE needs to drive MinD off the membrane. Hence,
\begin{equation}
	<\tau_{D}>=(\omega_{E}c_{E}+\omega_{ed}c_{e})^{-1}+\omega_{de}^{-1}.
\end{equation}
For the mean MinE residence time $<\tau_E>$ one has to consider the conditional branches whether MinE detaches alongside MinD (with probability $p_c$) or remains on the membrane (with probability $p_m=1-p_c$), and whether membrane-bound MinE detaches from the membrane (with probability $q_c=\omega_e/(\omega_e+\omega_{ed}c_d)$) or reassociates with MinD (with probability $q_m=1-q_c$):
\begin{equation}
	<\tau_{E}>=p_{c}\omega_{de,c}^{-1}+p_m(\omega_{de,m}^{-1}+q_{c}\omega_{e}^{-1}+q_m((\omega_{ed}c_{d})^{-1}+<\tau_{E}>)).
\end{equation}
After a few algebraic manipulations one obtains an expression for the mean MinE residence time
\begin{equation}
	<\tau_{E}>=\frac{2(\omega_{de,m}+\omega_{ed}c_{d}+\omega_{e})}{\omega_{de,c}\omega_{ed}c_{d}+\omega_{de}\omega_{e}},
\end{equation}
and for the ratio of the residence times $\Delta\tau$
\begin{equation}
	\Delta\tau=\frac{2\omega_{de}(\omega_{de,m}+\omega_{ed}c_{d}+\omega_{e})(\omega_{ed}c_e+\omega_{E}c_E)}{(\omega_{de,c}\omega_{ed}c_{d}+\omega_{de}\omega_{e})(\omega_{de}+\omega_{ed}c_e+\omega_{E}c_E)}.
\end{equation}
By comparison with the experiment $(\Delta\tau_\text{exp}\approx1.31)$ we obtain an expression relating all processes that characterize MinE interaction with the membrane to interactions with MinD. Due to the explicit dependency on the protein concentration $c_d$, $c_e$, and $c_E$ numerical simulations or additional approximations are required for further progression. We note that for the specific parameter choice in the present model $\Delta\tau$ becomes independent of $c_d$. The reason is that both exit processes for MinE occur on the same time scale, i.e. $\omega_{de,c}=\omega_e$, such that $<\tau_E>\rightarrow2/\omega_e$ and 
\begin{equation}
	\Delta\tau\rightarrow\frac{1}{p_c}\cdot\frac{2(\omega_{ed}c_e+\omega_{E}c_E)}{\omega_{de}+\omega_{ed}c_e+\omega_{E}c_E}.
\end{equation}
To obtain a first estimate we approximate the missing values for the protein concentrations by the stationary solution that represents the mean concentrations in linear approximation. For the parameters provided in the paper we find $\Delta\tau\approx14.7$, hence an 11-fold deviation from the experimental value $\Delta \tau_{\text{exp}}\approx 1.31$. We conclude that the model's assumption regarding MinE dynamics (in terms of parameter choice) clearly conflicts with experimental data. It represent the underlying molecular processes. In the following section we will investigate the model's linear stability upon varying the MinE dynamics.

\subsubsection{Linear Stability for varying MinE dynamics} 
\label{ssub:linear_stability_for_varying_mine_dynamics}
\begin{figure}[t]
	\centering
		\includegraphics[width=\textwidth]{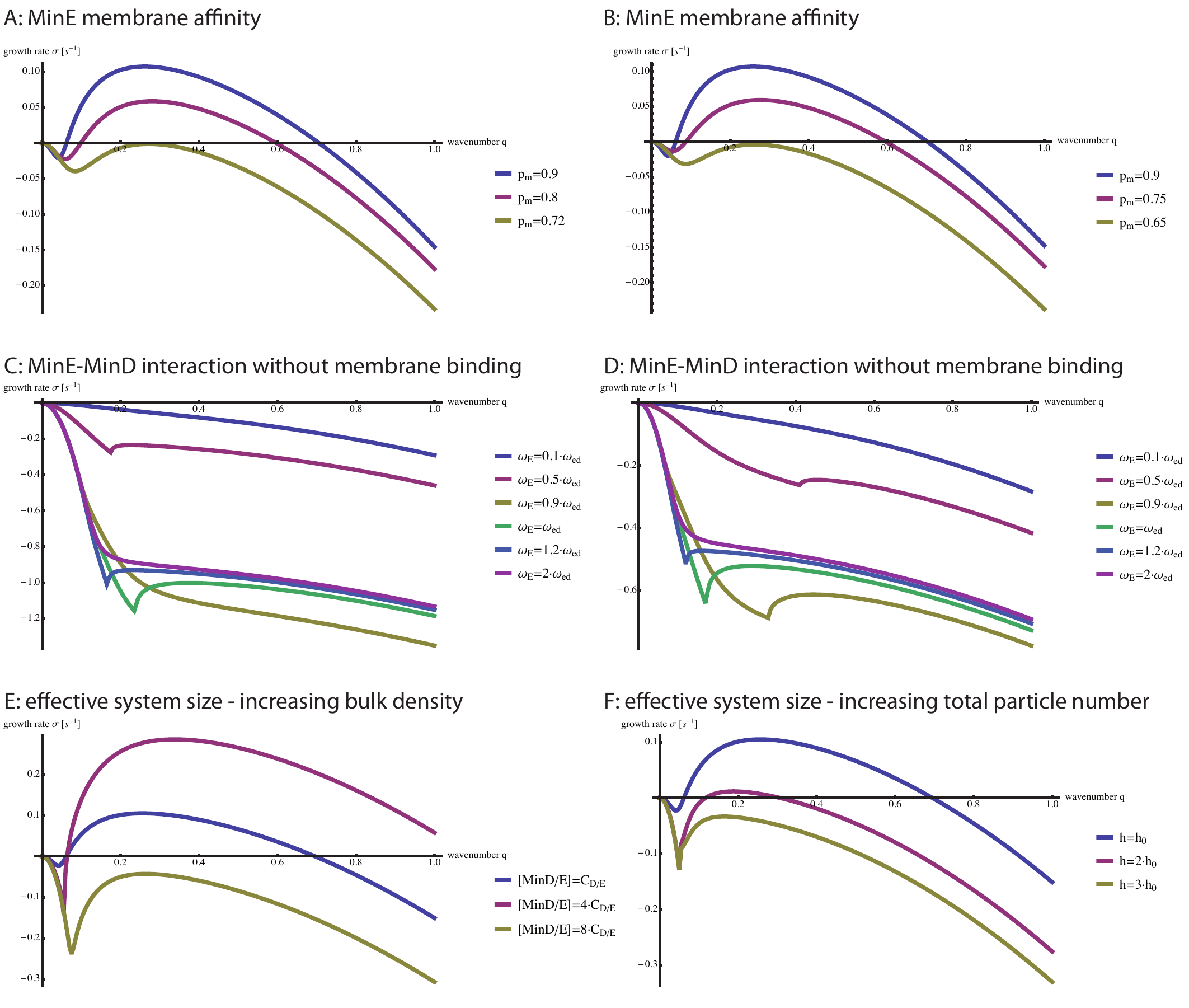}
	\caption{Dispersion relations for varying MinE-MinD interactions and effective system sizes showing loss of instabilities.}
	\label{fig:figures_linstab}
\end{figure}

The research by Park et al. \cite{Park2011} indicates that membrane associated MinE has an increased MinE-MinD interaction rate due to an exposed contact helix compared to the stable conformation in solution where the contact helix is buried in the dimeric interface. It is our understanding that this is the main idea which defines \emph{persistent binding} \cite{Loose2011a} and \emph{tarzan of the jungle} \cite{Park2011} models. To keep these mechanisms unaltered while trying to reduce the relative residence times we focus on the isolated effect of MinE membrane interactions. The membrane affinity of MinE can be expressed by $p_m=1-p_c$ alone. Tuning this parameter enables a smooth reduction of the relative residence time $\Delta \tau$ without altering the reassociation process with MinD. By decreasing the MinE membrane affinity we find that the dynamical instability is lost for $p_m<0.72$ at $\Delta \tau=3.9$, c.f. Figure~\ref{fig:figures_linstab}A. Releasing the constraint $\omega_{de,c}=\omega_e$ and increasing $\omega_e$ to weaken MinE membrane binding similarly leads to loss of dynamical instability for $\omega_e>0.65$ at $\Delta \tau=4.5$, c.f. Figure~\ref{fig:figures_linstab}B. We draw two conclusions from these results. First, tweaking the MinE membrane affinity alone proves to be insufficient to recover consistency with the experimental data. It appears that the problems are at least inherent to the parameter configuration as a whole (in particular, the MinE-MinD interactions) such that the model requires a broad reinvestigation in it's vast parameter space.\footnote{We note that reducing the MinE-MinD interaction parameters $\omega_E$ and $\omega_{ed}$ leads to loss of instability as well.} Second, the fact that the model requires at least 72\% of MinE remaining membrane bound after stimulating MinD ATPase activity implies that strong membrane interactions are a key requirement for pattern formation in the proposed model. Translated into an experimental test a hypothetical MinE mutant with slightly reduced membrane binding affinity but otherwise unaltered protein-protein interactions should lack any Min oscillations. In that case one would expect to observe the corresponding phenotypes (filamentous cells and minicells). However, we couldn't find any experimental evidence that supports this major model prediction. In Park et al. \cite{Park2011} four mutations of the MinE MTS ($\textit{minE}^{L3E}$, $\textit{minE}^{F7E}$, $\textit{minE}^{L4E}$, $\textit{minE}^{F6E}$) are investigated, quoting \cite{Park2011}:
\begin{quote}
	``Surprisingly, the strains containing $\textit{minE}^{L3E}$ and $\textit{minE}^{F7E}$ were extremely filamentous and could not
		form colonies on plates with arabinose (Figure 4B and data not shown), indicating that MinE function was absent. In contrast, strains containing $\textit{minE}^{L4E}$ and $\textit{minE}^{F6E}$ formed colonies normally on plates with arabinose, but the morphologies of the cells were heterogeneous in length with some minicells. The average cell length of an exponential culture of the strain with $\textit{minE}^{WT}$ was $2.84 \pm 0.89 \mu m$ compared to $4.68 \pm 2.48 \mu m$ for the strain lacking Min function. The strains containing $\textit{minE}^{F6E}$ and $\textit{minE}^{L4E}$ had average cell lengths of $3.81 \pm 2.67 \mu m$ and $2.95 \pm 1.37 \mu m$, respectively ($N \sim 250$ for each).
		In summary, each of the four charge substitution mutations eliminated membrane binding of the $\text{MinE}^{I25R}$ mutant. However, two of the mutations, $\textit{minE}^{L3E}$ and $\textit{minE}^{F7E}$, completely eliminated the ability of MinE to counteract MinC/MinD, whereas the other two, $\textit{minE}^{L4E}$ and $\textit{minE}^{F6E}$, did not, although they did reduce the ability of MinE to spatially regulate division as evidenced by the increases in the average cell length and the standard deviation.''
\end{quote}
Hence, removing MinE membrane binding affects the performance of the Min system rather than disabling it's function altogether. As discussed in our previous work \cite{Halatek:2012gm}, altering Min protein recruitment rates (which represent the sensing step discussed before) has significant impact on midcell localization by pole-to-pole oscillations. Therefore, the observed filamentous phenotypes in Park et al. \cite{Park2011} might be explained by an altered Min oscillation with reduced midcell localization precision due to weakened MinE-MinD interactions. The explanation offered by Park et al. \cite{Park2011} appears to agree with this line of thought, quoting \cite{Park2011}:
\begin{quote}
	`` [...] two of the MinE mutants we described above, $\text{MinE}^{F7E}$ and $\text{MinE}^{L3E}$, were unable to rescue cells from expression of MinC/ MinD (Figure 4B). This was surprising because these residues lie beyond the putative interacting helix. We reasoned that these residues could play a role in sensing MinD and therefore might have a defect in MinD-MinE interaction similar to that of the $\text{MinD}^{M193L}$ mutant. If so, the $\textit{minE}^{I24N}$ mutation should suppress these mutations. As shown in Figure 2D, the double mutant $\text{MinE}^{F7E/I24N}$ rescued cells from expression of MinC/MinD, demonstrating that the $\textit{minE}^{I24N}$ mutation is an intragenic suppressor of $\textit{minE}^{F7E}$. It also suppressed $\textit{minE}^{3LE}$ [\emph{sic}] (data not shown).''
\end{quote}
This picture is also consistent with the earlier experimental results by Ma et al. \cite{Ma2003} which showed that $\text{MinE}^{I25R}$ lacks the ability to sense MinD. In turn, this leads to the loss of Min-oscillations and results in the filamentous phenotype. The fact that $\textit{minE}^{L4E}$, $\textit{minE}^{F6E}$, and $\text{MinE}^{F7E/I24N}$ seems to retain the function of the Min system, while lacking the ability to bind to the membrane clearly implies that not membrane binding itself but the effect it has on MinD sensing enables proper Min oscillations. For $\textit{minE}^{L4E}$ and $\textit{minE}^{F6E}$ the specific effect on MinD sensing cannot be deduced from the experimental data. This impedes an unequivocal comparison with the model. 
However, this is not the case for $\text{MinE}^{F7E/I24N}$.  $\text{MinE}^{I24N}$ shows significantly increased MinD sensing ability through unmasking of the anti-MinCD domains ($\beta 1$ strands) which are buried in the dimeric interface in WT MinE. So, on one hand $\text{MinE}^{F7E/I24N}$ lacks the ability to interact with the membrane, such that membrane binding effects can be excluded. On the other hand the unmasked anti-MinCD domains enable it to interact with MinD as if it were in it's membrane bound conformation. As the experiment indicates that this mutation restores proper function of the Min system this scenario can be directly translated into a test of the model dynamics: We incorporate the corresponding modification by preventing MinE to bind to the membrane $\omega_{de,m}=0/s$, $\omega_{de,c}=0.88/s$ and setting the recruitment rate $\omega_E$ for cytosolic MinE equal to the reassociation rate  $\omega_{ed}$ of former membrane bound MinE species, i.e. $\omega_E=\omega_{ed}=2.5\cdot 10^{-3}\mu m^2/s$. The experimental observation implies that the dynamical instability should be restored. In contrast, the linear stability analysis reveals no instabilities in this kinetic configuration, cf. Figure~\ref{fig:figures_linstab}C.\footnote{Figure~\ref{fig:figures_linstab}D shows the same data as Figure~\ref{fig:figures_linstab}C but with the hydrolysis rate $\omega_{de}$ reduced by 50\%. This corresponds to the measurement of reduced MinD ATPase activity stimulation by $\text{MinE}^{I24N}\text{-h}^*$ in the experiment by Park et al. \cite{Park2011}.} Moreover, we find that without MinE membrane binding the instability cannot be restored by changing the MinE recruitment rate. Therefore, not the modified MinE-MinD interaction but MinE membrane binding is the crucial component in the model. Assuming that the $\text{MinE}^{F7E/I24N}$ mutant most likely restores pole-to-pole Min oscillations (hence, geometry sensing) without requiring membrane binding further questions the claim that MinE membrane binding is responsible for geometry sensing, we conclude that the experiments by Park et al. \cite{Park2011} disprove the proposed model. We stress that a loss of dynamical instability without MinE membrane binding does not imply that MinE membrane bindings is required or responsible for geometry sensing in any way. The demonstration is given by the model simulations above: There one observes pattern formation based on dynamical instabilities but no geometry sensing.



\subsection{Total particle numbers, bulk-membrane ratio, and effective 2D modeling} 
\label{ssub:total_particle_numbers_bulk_membrane_ratio_and_effective_2d_modeling}
In this last section we focus on questions about volume effects, spacial dimensions, and effective system size. A main claim of the theoretical investigation in the paper is that
\begin{quote}
``	[...] Min protein waves sense the geometry of the flat, two-dimensional membrane, rather than the three-dimensional space of the cell or the curvature of the membrane.
''\end{quote}
Certainly, there is no doubt that the available experimental data offers such a conjecture as waves are found for various system/bulk heights and the gold layer size does not seem to have any impact beyond enabling and disabling patch-to-patch coupling. However, in light of the numerical investigation above this statement raises the question why the size of the space above the membrane should not matter while additional space around the membrane (in form of the surrounding gold layer) has significant impact on the model dynamics. Increasing gold layer size leads to loss of patterns for the experimental [MinE]/[MinD] ratio and impedes wave alignment (i.e. geometry sensing phenomena) for fine tuned [MinE]/[MinD] ratios. This indicates that bulk size (via bulk/membrane ratio) affects wave dynamics. Furthermore, our investigation showed that alignment to rectangular membrane patches solely emanates from cross-boundary coupling effects. This directly contradicts the model-based claim that ``\emph{Min proteins waves sense the geometry of the flat, two-dimensional membrane}''. 
The foundation of this conclusion appear inscrutable to us. The authors state in the supplementary document to the paper that they ``\emph{have checked on specific examples that the same phenomena [...] can also be observed in the full three-dimensional description}''. Obviously this requires an explicit mapping between the full 3D dynamics and the effective 2D reduction. However, no such mapping is provided in the model definition. Inspection of the model parameters reveals that the cytosolic protein concentrations are treated as surface densities: $C_{D0}=2.9 \cdot 10^3 \mu m^{-2}$ and $C_{E0}=1.9 \cdot 10^3 \mu m^{-2}$. 
This indicates an underlying bulk integration. 
The protein concentrations in the experiment are $c_D=0.8\mu M\approx 481.8 \mu m^{-3}$ MinD and $c_E=0.5\mu M\approx 301.1 \mu m^{-3}$ MinE. Comparison with the model parameters implies integration of $6\mu m$ bulk, i.e. $C_{D0/E0}\approx 6 c_{D/E}$. 
The paper does not provide the experimental bulk height explicitly, however, it is stated that the bulk height is very large, quoting \cite{Schweizer:2012ba}:
\begin{quote}
``Although the space above the membrane was not limited in our experiments, the proteins were located only in a small layer above the membrane during pattern formation.''
\end{quote}
The corresponding figure (Figure S1 in the supplement of the paper) clearly shows that the experimental bulk height is much larger than $6\mu m$, and previous in vitro experiments were performed with a total bulk height about $h=5 \cdot 10^3\mu m$ \cite{Loose:2008ca}. Regarding particle numbers the model accounts for a system that is three orders of magnitude smaller than the typical experimental setup. 

Without any notion of bulk volume in the model definition we are left with ad hoc approximations that maintain the mathematical structure of the model. The bulk dynamics in normal direction to the membrane can be eliminated via integration or averaging to yield a 2D reaction-diffusion system. In the first case we increase the total densities of MinD $(C_{D0})$ and MinE $(C_{E0})$ keeping the ratio constant. In the second case we keep the total densities of MinD and MinE fixed and introduce a scaling factor $h$ between between membrane and bulk dynamics, i.e. 
\begin{equation}
	\partial_t c_B = D_B\nabla^2 c_B+\frac{1}{h}f_B
\end{equation} 
where $c_B$ denotes any bulk species with membrane reactions given by $f_B$. Note that in this case bulk densities are (mean) volume densities and not surface densities as the reported parameters suggest.
Using the effective system size as parameter we find that dynamical instabilities are rapidly lost in both approximation (Figure~\ref{fig:figures_linstab}E/F). This proves that the model's dynamics are actually highly sensitive to volume effects in contrast to the authors' claim in the paper.
We further note that these findings are supported by the fact that patterns vanish if the gold layer size is increased. 

We conclude that the model cannot account for experimental system sizes. Moreover there is no explicit notion of bulk volume and no relation to the full three-dimensional dynamics. Increasing the effective system size leads to loss of any pattern forming instabilities which directly contradicts the claim that three dimensional dynamics do not affect pattern formation. 
\newpage
\section{Figures and References} 
\label{sec:figures_and_references}


\begin{thebibliography}{20}

\bibitem{Arjunan2009}
A.~Arjunan and M.~Tomita.
\newblock {A new multicompartmental reaction-diffusion modeling method links
  transient membrane attachment of E. coli MinE to E-ring formation}.
\newblock {\em Syst. Synth. Biol.}, 4:35--53, 2009.

\bibitem{cross2009}
M.~Cross and H.~Greenside.
\newblock {\em {Pattern Formation an Dynamics in Nonequilibrium Systems}}.
\newblock Cambridge University Press, 2009.

\bibitem{Halatek:2012gm}
J.~Halatek and E.~Frey.
\newblock {Highly Canalized MinD Transfer and MinE Sequestration Explain the
  Origin of Robust MinCDE-Protein Dynamics}.
\newblock {\em Cell Reports}, 1(6):741--752, 2012.

\bibitem{Huang:2003bc}
K.~Huang, Y.~Meir, and N.~Wingreen
\newblock {Dynamic structures in Escherichia coli: Spontaneous formation of
  MinE rings and MinD polar zones}.
\newblock {\em Proc. Natl. Acad. Sci. USA}, 100(22):12724--12728, 2003.

\bibitem{Kerr2006}
R.~A. Kerr, H.~Levine, T.~J. Sejnowski, and W.-J. Rappel.
\newblock {Division accuracy in a stochastic model of Min oscillations in
  Escherichia coli.}
\newblock {\em Proc. Natl. Acad. Sci. USA}, 103(2):347--352, 2006.

\bibitem{Loose2011a}
M.~Loose, E.~Fischer-Friedrich, C.~Herold, K.~Kruse, and P.~Schwille.
\newblock {Min protein patterns emerge from rapid rebinding and membrane
  interaction of MinE}.
\newblock {\em Nat. Struct. Mol. Biol.}, 18(5):577--583, 2011.

\bibitem{Loose:2008ca}
M.~Loose, E.~Fischer-Friedrich, J.~Ries, K.~Kruse, and P.~Schwille.
\newblock {Spatial Regulators for Bacterial Cell Division Self-Organize into
  Surface Waves in Vitro}.
\newblock {\em Science}, 320(5877):789--792, 2008.

\bibitem{Ma2003}
L.~Ma, G.~King, and L.~Rothfield.
\newblock {Mapping the MinE site involved in interaction with the MinD division
  site selection protein of Escherichia coli}.
\newblock {\em J. Bacteriol.}, 185(16):4948--4955, 2003.

\bibitem{Meacci2005}
G.~Meacci and K.~Kruse.
\newblock {Min-oscillations in Escherichia coli induced by interactions of
  membrane-bound proteins.}
\newblock {\em Phys. Biol.}, 2(2):89--97, 2005.

\bibitem{Park2011}
K.-T. Park, W.~Wu, K.~P. Battaile, S.~Lovell, T.~Holyoak, and J.~Lutkenhaus.
\newblock {The Min Oscillator Uses MinD-Dependent Conformational Changes in
  MinE to Spatially Regulate Cytokinesis.}
\newblock {\em Cell}, 146(3):396--407, 2011.

\bibitem{Schweizer:2012ba}
J.~Schweizer, M.~Loose, M.~Bonny, K.~Kruse, I.~M{\"o}nch, and P.~Schwille.
\newblock {Geometry sensing by self-organized protein patterns.}
\newblock {\em Proc. Natl. Acad. Sci. USA}, 109(38):15283--15288, 2012.

\end{thebibliography}
\end{document}